\documentclass[10pt, format=manuscript,nonacm]{acmart}
\usepackage{float}
\usepackage{tabularx}  
\usepackage{pifont}
\usepackage{subfig}

\usepackage{url}
\AtBeginDocument{%
  \providecommand\BibTeX{{%
    \normalfont B\kern-0.5em{\scshape i\kern-0.25em b}\kern-0.8em\TeX}}}

\setcopyright{acmcopyright}
\copyrightyear{2018}
\acmYear{2018}
\acmDOI{XXXXXXX.XXXXXXX}

\acmConference[Conference acronym 'XX]{Make sure to enter the correct
  conference title from your rights confirmation emai}{June 03--05,
  2018}{Woodstock, NY}
\acmBooktitle{Woodstock '18: ACM Symposium on Neural Gaze Detection,
 June 03--05, 2018, Woodstock, NY} 
\acmPrice{15.00}
\acmISBN{978-1-4503-XXXX-X/18/06}

\begin{document}
\title{CogniDot: Vasoactivity-based Cognitive Load Monitoring with a Miniature On-skin Sensor}

\author{Hongbo Lan}
\affiliation{%
  \institution{Institute of computing technology, Chinese academy of sciences}
  \country{China}
}
\affiliation{%
  \institution{Tsinghua university}
  \country{China}
}
\email{lanhb.box@gmail.com}

\author{Yanrong Li}
\affiliation{%
  \institution{Chinese Academy of Sciences}
  \country{China}
}

\author{Shixuan Li}
\affiliation{%
  \institution{Tsinghua University}
  \country{China}
}

\author{Xing Yi}
\affiliation{%
  \institution{Institute for Network Sciences and Cyberspace, Tsinghua University}
  \country{China}
}

\author{Tengxiang Zhang}
\authornote{*}
\affiliation{%
  \institution{Institute of Computing Technology, Chinese Academy of Sciences}
  \country{China}
}
\newcommand*{\eg}{\textit{e.g.},\;}
\newcommand*{\ie}{\textit{i.e.},\;}
\newcommand*{\vs}{\textit{v.s.}\;}
\newcommand*{\etc}{\textit{etc.}}
\newcommand*{\st}{\textit{s.t.},\;}
\newcommand*{\etal}{\textit{et~al.}\;}
\newcommand\todoyr[1]{\textcolor{red}{\textit{TODO\_YANRONG: #1}}}
\newcommand\todosx[1]{\textcolor{orange}{\textit{TODO\_SHIXUAN: #1}}}
\newcommand\todohb[1]{\textcolor{purple}{\textit{TODO\_HONGBO: #1}}}
\newcommand\todo[1]{\textcolor{blue}{\textit{TODO: #1}}}

\newcommand{\code}[1]{\texttt{#1}}



\begin{abstract}
Vascular activities offer valuable signatures for psychological monitoring applications. We present CogniDot, an affordable, miniature skin sensor placed on the temporal area on the head that senses cognitive loads with a single-pixel color sensor. With its energy-efficient design, bio-compatible adhesive, and compact size (22mm diameter, 8.5mm thickness), it is ideal for long-term monitoring of mind status. We showed in detail the hardware design of our sensor. The user study results with 12 participants show that CogniDot can accurately differentiate between three levels of cognitive loads with a within-user accuracy of 97\%. We also discuss its potential for broader applications.
\end{abstract}



\maketitle
\section{Introduction}
Cognitive load is an important indicator for both holistic psychological health and task-specific mental effort. 
From a health perspective, tracking an individual's cognitive load can provide invaluable insights into mental well-being. 
For applications in human-computer interaction, quantified cognitive load allow evaluating user mental effort during interaction, enabling systems to dynamically adapt based on user cognitive states. 
For instance, an educational software could adjust its teaching pace according to the learner's cognitive load, ensuring efficient learning without overwhelming them. 
Similarly, driver assistance systems can gauge the cognitive load of the driver, intervening or alerting when cognitive overload suggests decreased attention. 

One common method to measure cognitive load involves the use of questionnaires and self-reported scales (\eg NASA-TLX form), where participants rate their perceived mental effort or challenge after a task is completed. 
Even though subjective forms can provide a quick snapshot of the cognitive state, they cannot support real-time cognitive load measurement. 
For real-time measurement of cognitive load, researchers look at various physiological signals that changes under different mental states. 
For example, by capturing subtle temperature changes on different facial areas like forehead and nose, researchers can determine whether the user has a high cognitive load by analysing the thermal patterns\cite{abdelrahman2017Cognitiveheat}.
The principle is that there are more blood flown to the brain during high cognitive load, leading to specific facial thermal patterns.  

In this paper, we propose to quantify cognitive loads by observing vessels. 
Vascular activities, ranging from rhythmic pulsations of blood vessels to the flow volume and composition of blood, can provide rich insight into various human states.  

Vascular signals can reveal changes in psychological states. 
For example, anxiety and depression can cause vessel dysfunction for their high responsiveness to autonomic and hormonal changes [association of anxiety, stillman].
In terms of cognitive loads, the brain has different energy and oxygen demands, leading to fluctuations in blood flow, as well as the level of blood oxygen \cite{FunctionalOpticalBrain}. Vascular sensors can directly capture subtle changes in blood flow patterns and compositions, eliminating the need for expensive and high-power thermal cameras that indirectly sense blood flow rate increases by measuring temperature. 
However, traditional vascular sensing equipment like blood meters\cite{wyatt1968electromagnetic,hoskins1990ultrasound} are usually bulky and expensive, not suitable for everyday use. 
Photoplethysmograms (PPGs) sensors and oximeters are widely wearable vascular sensors. 
They measures the relative changes in blood volume or blood oxygen saturation by shining lights that penetrate the skin into a blood vessel, then measuring the amount of light reflected to photodiodes. 
However, such sensors are mostly integrated with smart wearables like smartwatches. 
The limited sensing position on-body greatly constrains the scope of applications for such sensors. 

We design CogniDot to address the challenge for a low-cost and easy-to-use sensor that can be placed on the head for vascular activity monitoring. 
CogniDot is a miniature skin sensor that can be easily placed on the head, or anywhere on the body, for nearly 8 hours of non-stop vasoactivity monitoring.
The sensing dot has a small diameter of 22mm, a low-profile of 8.5mm, and a lightweight of 6.45g.
The bio-compatible adhesive gel of CogniDot enables fast and easy attachment on the temporal area on the face, where vessels responsible for major blood flow into and out of the brain locate\cite{kleintjes2007forehead}.
CogniDot uses a single-pixel color sensor TCS3430 (AMS Inc.) that covers both visible and near-infrared (NIR) spectrum, which enables a holistic view of vascular dynamics. 
The reflected light energies in three visible and two NIR channels are synthesized to reveal activities of under-skin vessels and capillaries in a non-invasive and efficient manner. 

CogniDot has several advantages compared with existing cognitive load detection solutions. 
Firstly, instead of integrated into existing smart wearables, CogniDot is a much more flexible standalone vascular activity sensor;
Secondly, the miniature size and lightweight makes it negligible for the user when placed on the head. 
Together with the low power consumption, CogniDot support comfort long-term wearing for continuous monitoring of the cognitive load;
Finally, the wide spectrum covering both visible and NIR bands offers rich signal features that improves the overall cognitive load detection performance. 

We conducted a user study to validate the ability of CogniDot to accurately quantify different levels of cognitive loads. 
The results show that CogniDot can accurately differentiate between three levels of cognitive loads with a within-user accuracy of 97\%.

Our contribution is three-fold:

\begin{itemize}
    \item We build a miniature and low-cost sensor that can be easily attached around the head temporal area for long-term brain-related vasoactivity monitoring.
    \item We show that visible light can be used to calibrate the NIR data drift due to sensor-to-skin distance and ambient light strength changes. 
    \item We develop a vascular signal processing pipeline that accurately differentiate between three levels of cognitive loads.
\end{itemize}

\section{Background and Related Work}
In this section, we review psychological monitoring research work that relies on vascular activities, and existing solutions for wearable vasoactivity sensing.  

\subsection{Vasoactivity-based Psychological Sensing}

Vascular activity includes factors such as blood flow, blood perfusion rate, blood cell composition, blood pressure, as well as the dilation and constriction of blood vessels.
It is closely related with many human psychological processes.
As biological indications, vascular activity reveals how the body and the brain respond and adapt to various challenges.

Vascular activity can reflect a person's emotional and cognitive states. 
Blood perfusion rate can be used to assess the blood supply to the brain, reflecting the activity in different regions of the brain\cite{doi:10.1038/sj.jcbfm.9600526}. 
Z.Zhu \etal~\cite{zhu2007forehead} proved that when people lie, their supraorbital vasculature blood perfusion rate will increase, leading to a rise in forehead skin thermal radiation, a similar phenomenon will also occur in the periorbital region\cite{Wang_Leedham_2006}. 
In addition, some diseases such as schizophrenia\cite{liddle1992cerebral} and depression\cite{cantisani2016distinct} can also be reflected in abnormal changes in cerebral blood flow perfusion. 
Matsukawa \etal~\cite{Matsukawa_Endo_Ishii_Ito_Liang_2018} proved that emotions of anger and anxiety would lead to vasoconstriction, while relaxation would lead to vasodilation.

Facial temperature measurement is the most commonly used method for cognitive load monitoring.
Abdelrahman~\etal \cite{Abdelrahman_Velloso_Dingler_Schmidt_Vetere_2017} employed a commercial high-resolution thermal camera to observe temperature variations on participants' noses and foreheads during text readings and Stroop tests. 
They found that as cognitive load intensified, forehead temperature increased and nose temperature decreased, widening the temperature gap between the two. 
Captivates \cite{Chwalek_Ramsay_Paradiso_2021} utilized compact thermoelectric sensors on eyeglasses to measure this temperature disparity, confirming Abdelrahman's findings. 
Other than temperature-based techniques, Bonetti \etal proved when people's cognitive load increased, the content of oxygenated hemoglobin in their brain would also increase, and would return to the normal levels during rest\cite{doi:10.1038/sj.jcbfm.9600526}. 
Compared to temperature-based method, blood composition changes are more direct indicators of the brain status. 

In this work, we also monitor oxyhemoglobin rate changes for cognitive load tracking.   
Instead of measuring blood compositions in the brain area, we place our sensor on the temporal area to monitor the vascular activities of superficial temporal arteries, which are responsible for brain oxygen and energy supply. 
In this way, we eliminate the need for advanced sensing equipment required for brain vessel sensing, enabling low-cost and long-term cognitive load monitoring.

\subsection{Existing Vasoactivity Sensors}

Researchers have explored many sensing modalities for vascular activity sensing. In clinical diagnosis and biomedical research, some researchers use electromagnetic\cite{wyatt1968electromagnetic}, laser\cite{qu2014laserDoppler}, magnetic resonance\cite{battocletti1972nuclearmagnetics} or ultrasonic blood flow meter\cite{hoskins1990ultrasound} to detect vascular blood flow. 
Such sensing equipment is usually bulky and expensive, not suitable for everyday monitoring. For example, assessing a person's anxiety requires imaging the blood flow in various regions of the brain with magnetic resonance imaging (MRI)\cite{Alexopoulos_Meyers_Young_Kakuma_Silbersweig_Charlson_1997}. 

Wang~\etal~\cite{wang2018seismo} rely on computer vision techniques to sense facial vascular activities by analyzing the color composition changes over a certain period. 
In addition to smartphones, many commercial instruments and wearable devices can also monitor vascular signals through built-in sensors. 
For example, Ahemed~\etal~\cite{Ahmed_Banerjee_Ghose_Sinharay_2015} designed a pair of smart glasses to monitor users' heart rate and blood oxygen levels, ensuring better contact with skin and reducing artifact.
Webb~\etal~\cite{doi:10.1126/sciadv.1500701} designed an ultra-thin, flexible sensor that could adhere to the surface of the skin, enabling precise assessment of blood flow in both large and micro vessels.

Compared with existing solutions, CogniDot support continuous and long-term monitoring of cognitive load thanks to its comfortable wearing and low power consumption. 
It leverages both the visible and the NIR spectrum for vascular activity sensing of both superficial capillaries and under-skin vessels. 
In this way, CogniDot provides a more thorough feature set of activities of different types of vessels, enabling more accurate vasoactivity sensing. 

\section{Method}

\subsection{Sensing Principle}

Cognitive load refers to the amount of information or mental effort that the brain needs to handle while undertaking a cognitive task.  The capacity of an individual's working memory and the ability to process information is limited. When people are completing tasks of varying difficulty, their brains will experience different burdens.
Hundreds of studies in physiology and psychology have shown that cognitive load is an essential factor in sensing and understanding user behavior. 

When users are suffering from high cognitive load, their Physiological parameters will change consequently, such as the heart rate, blood pressure, sweating, and the change of blood components in vascular in the cerebral cortex. As cognitive load increases, there would be more flood flow into the brain\cite{zhu_forehead_2007}, and this also can be observed in the bilateral forehead zone \cite{liu_transdermal_2018}. 

We take near-infrared spectroscopy (NIRS) as our basic principle in measuring the concentrations of oxygenated hemoglobin (HbO2) and deoxygenated hemoglobin (Hb) in tissue vascular. In NIRS, lights in the wavelength range of 650-950 nm are used to illuminate tissue and then their reflection by the vascular through tissue is recorded to evaluate hemoglobin concentrations\cite{pinti_present_2020}. In our system, we focus on two critical wavelengths: 730 nm and 940 nm. It is noteworthy that deoxygenated hemoglobin (Hb) and oxygenated hemoglobin (HbO2) demonstrate distinct absorption characteristics at these respective wavelengths. So according to modified Beer-Lambert Law\cite{maki_spatial_1995}, we can calculate the concentration value of HbO2 and Hb: 

\begin{align}
C_{HbO_2} = \frac{\varepsilon_{\text{Hb}}^{\lambda_2} D^{\lambda_1} - \varepsilon_{\text{Hb}}^{\lambda_1} D^{\lambda_2}}{L (\varepsilon_{\text{Hb}}^{\lambda_2} \varepsilon_{HbO_2}^{\lambda_1} - \varepsilon_{\text{Hb}}^{\lambda_1} \varepsilon_{HbO_2}^{\lambda_2})}
\end{align}

\begin{align}
C_{Hb} = \frac{\varepsilon_{HbO_2}^{\lambda_2} D^{\lambda_1} - \varepsilon_{HbO_2}^{\lambda_1} D^{\lambda_2}}{L (\varepsilon_{\text{Hb}}^{\lambda_1} \varepsilon_{HbO_2}^{\lambda_2} - \varepsilon_{\text{Hb}}^{\lambda_2} \varepsilon_{HbO_2}^{\lambda_1})}
\end{align}

where $\varepsilon_{HbO_2}^{\lambda_i}$ and $\varepsilon_{Hb}^{\lambda_i}$ are the extinction coefficients of oxygenated and deoxygenated hemoglobin at wavelength ${\lambda_i}$. $D^{\lambda_i}$ is the optical density of reflected light at wavelength ${\lambda_i}$ received by the sensor. L is the light path, which is around 0.75cm in our system. Also, the changes in  $C_{HbO_2}$ and $C_{Hb}$ can be consequently determined:

\begin{align}
\varDelta C_{HbO_2} = \frac{\varepsilon_{\text{Hb}}^{\lambda_2} \varDelta D^{\lambda_1} - \varepsilon_{\text{Hb}}^{\lambda_1} \varDelta D^{\lambda_2}}{L (\varepsilon_{\text{Hb}}^{\lambda_2} \varepsilon_{HbO_2}^{\lambda_1} - \varepsilon_{\text{Hb}}^{\lambda_1} \varepsilon_{HbO_2}^{\lambda_2})}
\end{align}

\begin{align}
\Delta C_{Hb} = \frac{\varepsilon_{HbO_2}^{\lambda_2} \varDelta D^{\lambda_1} - \varepsilon_{HbO_2}^{\lambda_1} \varDelta D^{\lambda_2}}{L (\varepsilon_{\text{Hb}}^{\lambda_1} \varepsilon_{HbO_2}^{\lambda_2} - \varepsilon_{\text{Hb}}^{\lambda_2} \varepsilon_{HbO_2}^{\lambda_1})}
\end{align}

where  $\varDelta D^{\lambda_i}$  is the change in optical density at wavelength ${\lambda_i}$ and can be calculated by the following equation:
\begin{align}
    \varDelta D^{\lambda _{i}} = log {\frac{I_B}{I_T}}
\end{align}
where $I_B$ and $I_T$ are the light intensities at baseline and time T measured by the sensor, which are also the values read from IR1 and IR2 channel of our system.

\subsection{Hardware Design and Implementation}
\begin{figure}
    \centering
    \includegraphics[scale=0.4]{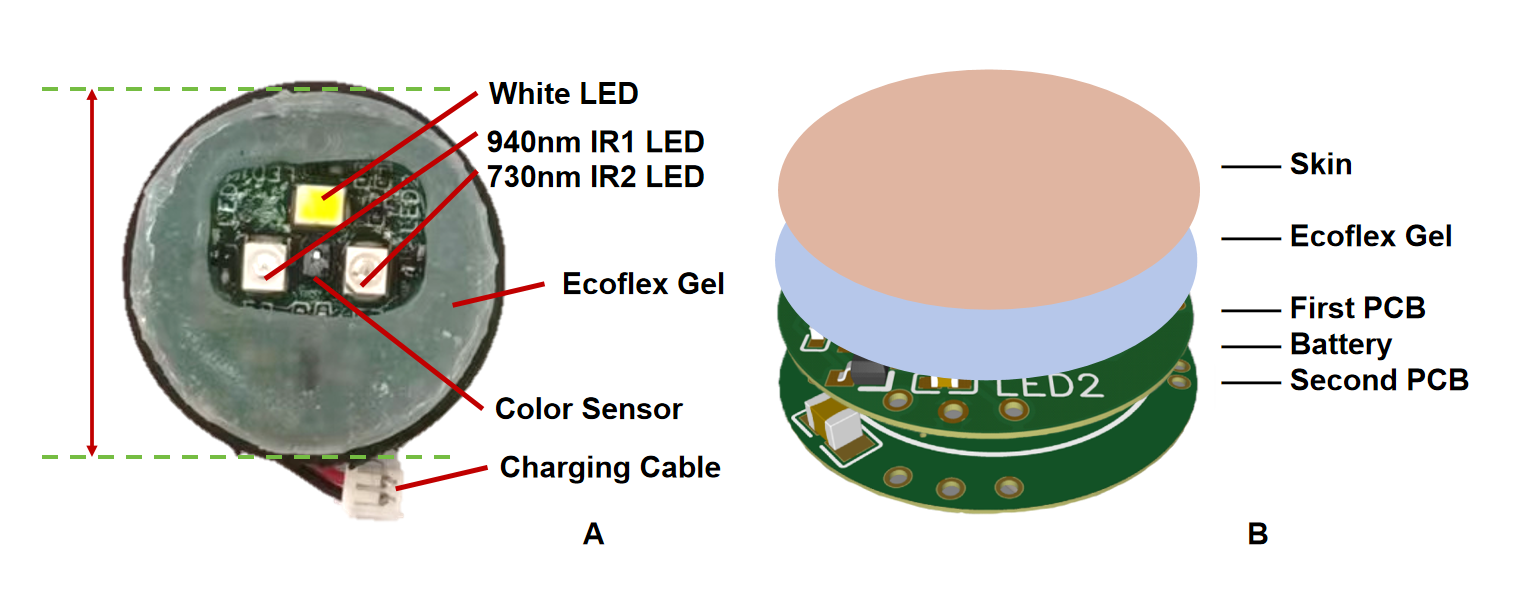}
    \caption{\textbf{CogniDot Hardware.}}
    \label{fig:hardware}
\end{figure}

We design CogniDot to have a small size, low profile, and light weight so that it can either be used alone or be integrated into wearables like hats and headbands.
Our sensor takes a round shape with a diameter of 22mm and a height of 8.5mm with a weight of only 6.45g (Figure~\ref{fig:hardware}A). 
The skin-friendly adhesive gel allows repeated attachments onto the human skin for measurements.

CogniDot follows a sandwich structure involving two PCB boards with a coin battery in the middle as shown in Figure~\ref{fig:hardware}B. 
The sensing circuits resides on the top layer of the first PCB board that is closest to the skin.
It contains a five-channel TCS3430 color sensor (AMS Inc.) to capture reflected light from the skin and the vessel. 
There are also three light sources: a white LED for visible light, a 730nm infrared LED for IR1 band as specified by the sensor, and a 940nm infrared LED  for the IR2 band. 
In this way, we have a more flexible control of the lighting and sensing sequences for different bands, enabling different firmware to optimize for sensing performance and battery life. 
The bottom layer of the first PCB board and the top layer of the second PCB board serve as battery positive and negative electrodes, which can hold coin batteris with diameters of 16mm and height of 3.2mm. 
The bottom layer of the second PCB board has an nRF52832 Bluetooth module (9.4mm x 9.25mm) for transmitting sensor data to computing devices. 
Such a structure reduces the overall sensor weight to 6.45g(including 3D printed housing).
\subsection{Hardware Design and Implementation}
\begin{figure}
    \centering
    \includegraphics[scale=0.4]{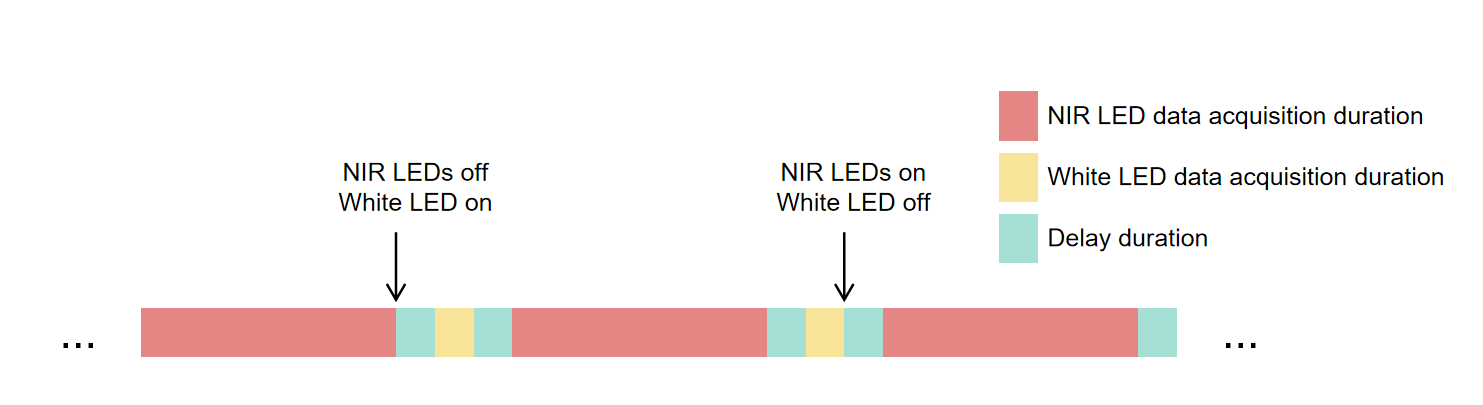}
    \caption{\textbf{LED control flow.}}
    \label{fig:LED}
\end{figure}
The sensor has to be placed on the skin in an easy way with robust adhesiveness. 
So, we developed the skin-friendly adhesive gel by combining Ecoflex-Gel-2 glue (AB liquid) with a C6H11BF4N2 solution in a weight ratio of A:C6H11BF4N2:B = 1:0.03:1. 
This mixture is poured into a 3D printed mold, resulting in a 3mm thick adhesive gel layer on the top layer of the first PCB board.
Not only is this gel harmless when in direct contact with the skin, but its adhesive properties also remain consistent even after repeated use, offering an advantage over traditional solutions like medical non-woven tapes. Overall, the power consumption is 31.635 mW, which is very low compared to other cognitive load sensing devices.

The hardware programming operates in cycles, where initially, the white light is activated to collect a single set of data. Following this, the white LED is turned off and the NIRs LED is switched on to gather data seven consecutive times before it is also turned off, marking the end of one cycle.Since the sensor is sensitive to color, in order to reduce the error effect caused by the alternation of light and dark, we implement a 100ms delay between the light turning on and the data collection to ensure the stability of the data( (Figure~\ref{fig:LED}).

\begin{figure}[h]
    \centering
    \includegraphics[width=\textwidth]{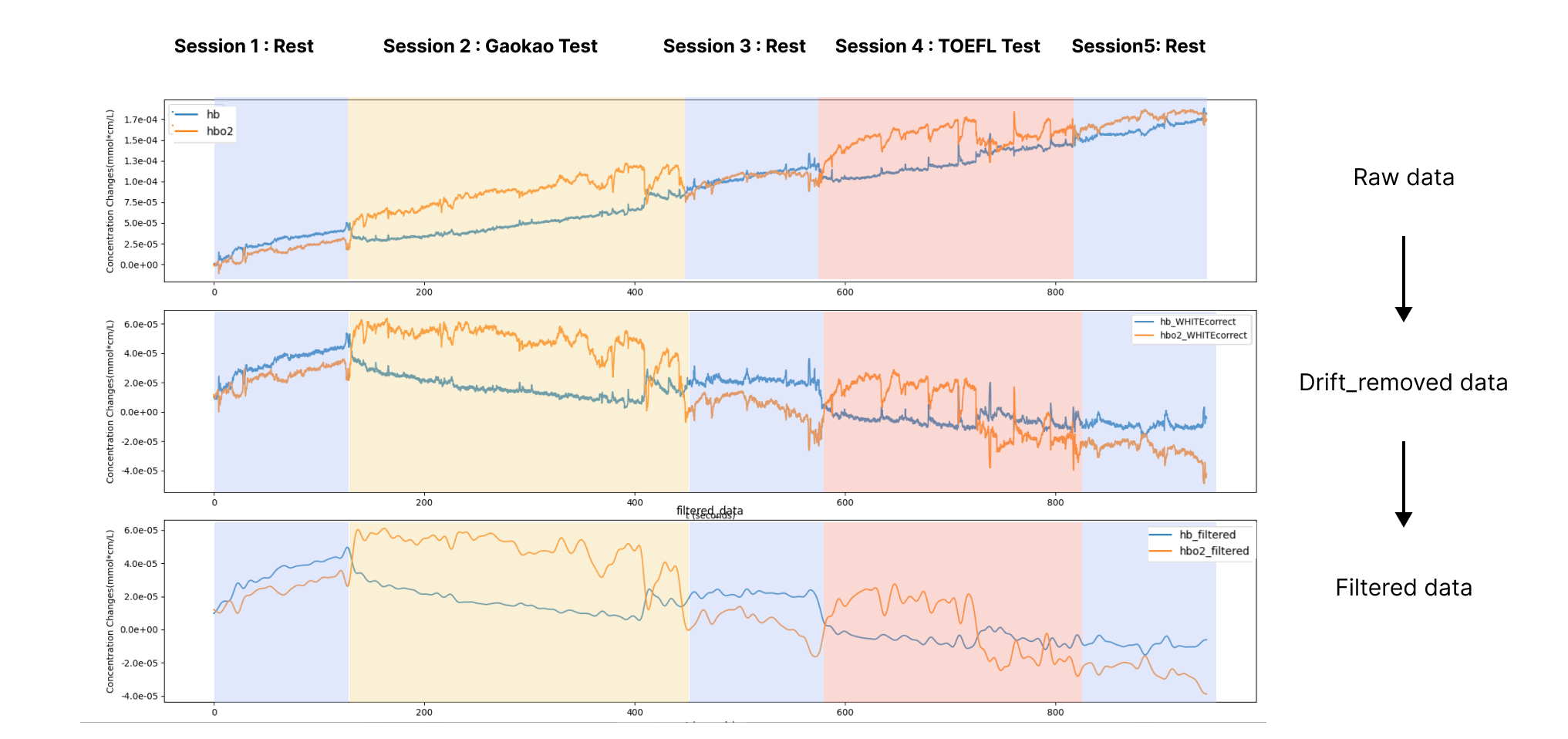}
    \caption{\textbf{Data processing of hemogoblin value and corresponding task session} - Distribution of Participant Data}
    \label{fig:wave}
\end{figure}

\subsection{Software Pipeline}

\textbf{Feature of HbO2/Hb}
TCS3430 has three color channels XYZ and two NIR channels (730nm and 940nm). We use the NIR channel data to calculate HbO2 and hb values. When a person's cognitive load is relatively high, HbO2 will increase, blood flow will increase, and Hb will decrease relatively. When a person returns to a normal resting state, Hbo2 will decrease and Hb will increase accordingly(Figure~\ref{fig:wave}).

\textbf{Removal of resting drift}

Besides, to better overcome the static drift phenomenon of sensors, we adopt the color sensor to correct our data by fitting the decay curve of channel Y, which represents the light intensity measured by our sensor. Since the lighting conditions in the confined space between the sensor and the skin are nearly static, it can be considered that the changing trend of the Y channel data is caused by static drift. 
A white LED flashes at a frequency of once per second, and the color sensor samples data whenever it illuminates. In other words, data sampling rate of the color sensor is 1 Hz. We utilize linear interpolation to fill and align the Y value sequence $Y_{raw}$, based on the length of the data sequence from the infrared channel. The obtained interpolated sequence $Y_{interp}$ is then subjected to regression fitting with the data from the infrared channel $IR_i$. The resulting residual values between them represent the actual data from the infrared channel after eliminating drift.

\textbf{Machine learning}
We performed a low-pass filtering with a cutoff frequency of 0.1 Hz on the signal after removing artifacts. Subsequently, we adopted a sliding window approach for data extraction. After multiple rounds of optimization and testing, we settled on using a 10-second window with a 5-second overlap, which means the window size is 70 frames, and the step size is 35 frames.

The final extracted data comprises the intensity values $IR_1$ and $IR_2$ of two infrared channels, along with the corresponding changes in hemoglobin concentrations $\Delta C_{HbO2}$ and deoxyhemoglobin concentrations $\Delta C_{Hb}$ at each time point. Subsequently, we conducted feature extraction from the data within each window by tsfresh library. Following this, we conducted machine learning algorithm analysis and sequentially verified the cognitive state classification results for each user.

\section{Evaluation}
We conducted a user study with cognitive tasks to measure the ability and user experience of our sensors in real-world scenarios, and probe the potential design implications for future work. 

\subsection{Experiment Design}
To measure the presence of our sensors in estimating three levels of cognitive load (rest, low, and high), we design an experiment flow with the following materials: English listening comprehension section in Chinese College Entrance Examination (Gaokao) and Test of English as a Foreign Language (TOEFL). The difficulty of TOEFL are obviously higher than Gaokao -- the latter produces much more cognitive load for two reasons: 1) it does not show the questions until finished playing the whole recording, 2) the recording only plays once in TOEFL while twice in Gaokao. Thus, participants have to memorize the content as much as possible while they are comprehending it. These two experimental materials correspond to two cognitive tasks with low and high cognitive load, respectively. 

The complete task flow is separated into five sessions: 2 minutes of rest, followed by 5 minutes of Gaokao listening comprehension section (consisting of two dialogues with a total of 6 questions), then another 2 minutes of rest, followed by 5 minutes of TOEFL listening comprehension section (featuring one dialogue with a total of 5 questions).

\begin{figure}
    \centering
    \includegraphics[width=0.4\textwidth]{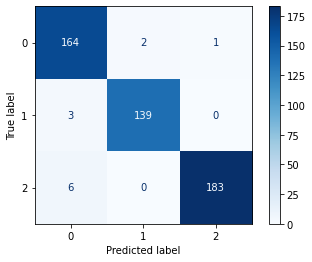}
    \caption{\textbf{Average confusion matrix of all participants.} - 0 represents rest state, 1 represents in low cognitive load stage and 2 represents in high cognitive load stage}
\end{figure}

\subsection{Participants and Questionnaire}
We recruited 12 participants (6 females, 6 males, aged 22-29) with good health conditions from the online community by posting recruitment messages. Considering that our cognitive tasks are English listening, we only recruited those who have basic English listening ability. Each participant are compensated with 12 US dollars. Informed consent was obtained from each participant before experiment started. 

A subjective questionnaire was delivered to the participants, aiming to collect their subjective evaluations of their English listening comprehension levels (rated on a scale of 1-5, with 1 indicating very poor, and 5 indicating very good, denoting the ability to understand the daily conversations held by native speakers). Moreover, the questionnaire sought to gauge participants' subjective perceptions of the cognitive load and difficulty level of different task sessions (rated on a scale of 1-5, with 1 indicating very easy, and 5 indicating very difficult). Participants were also invited to assess the portability and wearing comfort of the sensors.

\subsection{Procedure}
The experimental environment is equipped with the furniture of appropriate height and essential apparatus including patch sensors and two laptops; one for connecting to the main computer to collect and store data, another for participants to conduct experimental tasks, including read and answer the questions. 

Before the commencement of the experiment, we meticulously briefed the participants on the experimental procedure and the requirements of the TOEFL listening test to ensure they were familiar with the operation in the TOEFL mock test website\footnote{TOEFL Listening Practice: https://toefl.kmf.com/listening/pre?id=11mc6j}. The experiment is structured into five sessions. To begin with the first resting session, participants are instructed to relax with closed eyes, lean against the chair back, and abstain from moving or speaking for a duration of 2 minutes. This is followed by the first cognitive task where participants are asked to open their eyes and engage in a 5-minute Gaokao listening test. Subsequently, a second 2-minute resting session is initiated, replicating the conditions of the first resting session while preparing for the next session of the experiment. The second cognitive task then ensues, requiring participants to open their eyes, operate the mouse, and utilize paper for note-taking and answering questions over a span of 5 minutes. Concluding the task, a final resting session is introduced, mirroring the previous resting session for another 2 minutes before the data collection is terminated. If the participant completes the test in advance, we jump into the next session immediately but guarantee a complete rest period to ensure the participant's cognitive capability comes back to normal level. Upon the completion of data gathering, participants are requested to fill out the subjective questionnaire to collect individual assessments and feedback as mentioned above.

\subsection{Result Analysis}

We extracted light intensity data from two infrared channels from the sensors placed on the temporal area of each participant. These data were then inputted into a low-pass filter with a cutoff frequency of 0.1Hz for filtration. Using the aforementioned formula, we computed the concentrations of oxyhemoglobin and deoxyhemoglobin relative to their initial states, forming a four-channel data frame comprising (IR1, IR2, HbO2, Hb). Subsequently, we performed sliding segmentation with a time window of 10 seconds, acquiring sub-data series sets for each user. To evaluate the detection capability of our sensors in identifying the cognitive load of participants under the two cognitive stages of the experiment, we processed these data sets using classification techniques such as Random Forest models. The within-user results show that our average accuracy is 97.34\% for all participants.

\begin{table}[h!]
  \centering
  \caption{Classification comparison}
  \label{tab:example}
  \begin{tabularx}{0.9\textwidth}{|X|c|c|c|c|c|c|}
    \hline
    Samples & \multicolumn{3}{c|}{Three classification} & \multicolumn{3}{c|}{Binary classification} \\
    \hline
    \multicolumn{1}{|c|}{Accuracy/\%} & left temple & right temple & both temples &  left temple & right temple & both temples \\
    \hline
    \multicolumn{1}{|c|}{P1} & 38.1 & 38.5 & 38.3 & 61.8 & 60.5 & 61.4 \\
    \multicolumn{1}{|c|}{P2} & 39.9 & 39.6 & 39.7 & 62.5 & 60.4 & 60.7 \\
    \multicolumn{1}{|c|}{P3} & 38.1 & 37.9 & 38.0 & 61.9 & 62.1 & 62.2  \\
    \multicolumn{1}{|c|}{P4} & 37.9 & 37.6 & 37.8 & 62.0 & 62.4 & 62.5 \\
    \multicolumn{1}{|c|}{P5} & 26.3 & 26.6 & 26.4 & 73.7 & 73.4 & 73.3\\
    \multicolumn{1}{|c|}{P6} & 30.2 & 30.5 & 30.3 & 62.3 & 62.0 & 62.1 \\
    \multicolumn{1}{|c|}{P7} & 38.2 & 38.5 & 38.4 & 61.8 & 61.6 & 61.6  \\
    \multicolumn{1}{|c|}{P8} & 37.3 & 37.4 & 37.4 & 62.6 & 62.5 & 61.8  \\
    \hline
    \multicolumn{1}{|c|}{Average} & 35.6 & 35.9 & 36.0 & 62.6 & 63.1 & 63.2  \\
    \hline
  \end{tabularx}
\end{table}

According to the statistics from the questionnaires filled out by the participants, the distribution of English proficiency levels and the self-assessed disparity in difficulty levels of the two cognitive tasks (rated between 1-5, with higher values indicating greater discrepancy in difficulty levels) are illustrated in the subsequent pie chart and bar graph. Regarding the sensor's portability and wearing burden, two participants provided generally satisfactory feedback. However, they both expressed concerns about the potential risk of the sensor falling off, which could impact the user experience. They were not particularly pleased with the simplistic shape design of the sensor demo, believing that it could detract from their appearance when worn. One of the participants specifically mentioned being aware of the sensor's weight, without experiencing additional burden. Aside from these concerns, all other participants gave satisfactory evaluations.

\begin{figure}
    \centering
    \includegraphics[width=\textwidth]{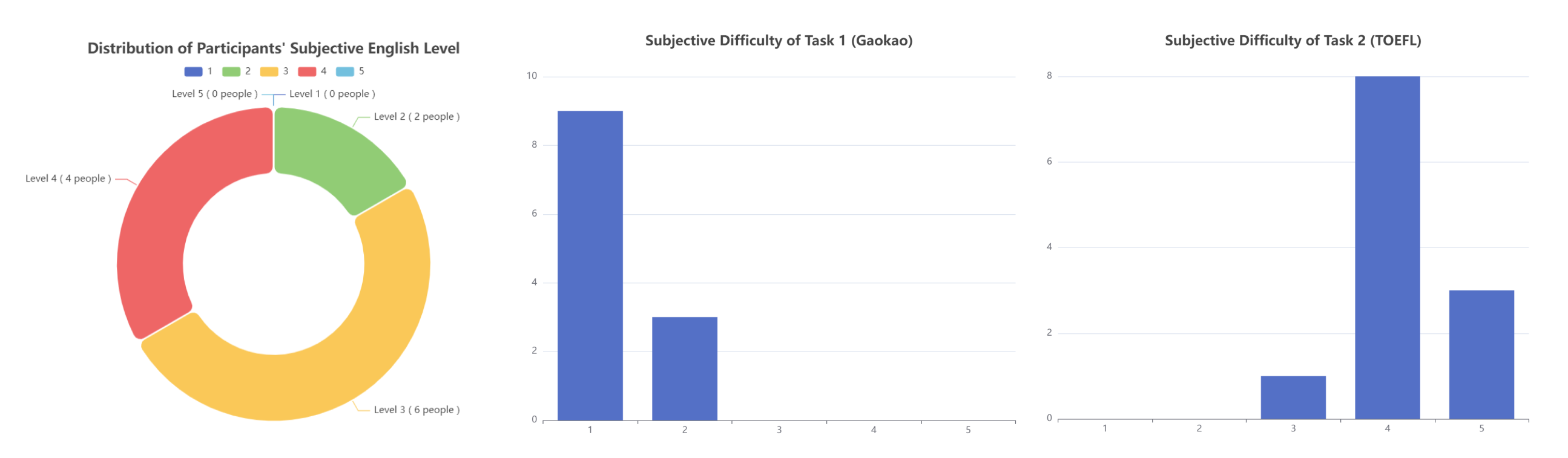}
    \caption{\textbf{Distribution of Participants' Subjective Scoring}. Left: bar charts represent the distribution of participants' subjective English proficiency levels. Right: bar graphs depict participants' subjective difficulty levels of Gaokao and TOEFL.}
\end{figure}

\section{Discussions}

\subsection{Feasibility of cognitive load detection using color sensor}
Cognitive load refers to the degree of brain power consumed by an individual's working memory during cognitive activities. Historically, most studies have relied on invasive techniques such as electroencephalography (EEG) \cite{antonenko2010electroencephalography}to measure cognitive load, which can detect subtle fluctuations in momentary load but can interfere with a subject's natural behavior and psychological state.In recent years, there has been a shift toward non-invasive techniques for monitoring cognitive load. One method involves monitoring pupil dilation\cite{wangwiwattana2018pupilnet}, a physiological response related to levels of cognitive load. In addition, Cognitive Heat\cite{abdelrahman2017Cognitiveheat} also proposed a method based on thermal imaging to detect cognitive load, trying to detect cognitive load unobtrusively.

These advances highlight the potential role of color sensors in detecting subtle physiological changes that occur during varying cognitive loads. However, the widespread use of cameras and thermal imaging cameras powered by large amounts of energy brings significant limitations, as they cannot be attached to the user for continuous monitoring due to their high power consumption.
Therefore, we believe that there are significant advantages in using lightweight, low-power sensors to detect cognitive load. This type of sensor can be seamlessly integrated into users' daily lives, providing an accurate and efficient means for real-time monitoring.

\subsection{Applications and Design Implications}
As hardware devices become more intelligent, cognitive load becomes particularly important as a means of perceiving human status. On current wearable devices, various physiological parameters of the user are detected to help optimize the user's daily activities and health management, including detecting cognitive load. By detecting cognitive load, smart home, industrial automation and other systems can automatically change strategies according to the user's status to reduce the user's cognitive load and improve their quality of life.
Since we noticed that there are currently few low-power modules specifically for cognitive load detection on the market, we designed a smart module with a small diameter of 22mm, a low-profile of 8.5mm. This miniaturized design is not only easy to embed into various wearable smart devices, such as headbands, hats, clothes, etc., but also its Bluetooth transmission characteristics provide the possibility of simultaneous detection of multiple modules, which means that we can Point detection to discover more fine-grained cognitive load conditions.

\section{Limitation and Future Work}
Although the gel sticks to the arm and will not fall off when you swing it, as time goes by, our skin secretions will still cause its stickiness to decrease and adhere to the gel. However, the gel is removable and will not damage the overall device. During previous experiments, we also replaced the gel to ensure the normal progress of the experiment. We then plan to design it into the shape of a suction cup based on the gel to further improve its adhesion to the skin.
CogniDot is currently equipped with a single-pixel color sensor. There may be a position where the sensor cannot directly align with the blood vessel. However, experiments have proven that single-pixel can still achieve basic cognitive load processes. We hope that in the future, based on the original specifications, we will replace it with smaller-scale LEDs, optimize the overall hardware structure, and equip it with an array-type sensor layout to achieve homologous multi-point measurements in the same module. Also, we will continue to explore more applications related to muscles and heart rate.
\section{Conclusion}
In this paper, we introduce CogniDot, a low-power, low-cost, and compact (22mm diameter, 8.5mm thickness) cognitive load detection module. We designed a five-stage cognitive task to evaluate the cognitive load detection ability of our sensor on 12 participants. The results showed that the average accuracy of the three classifications of within-user cognitive load among 12 users was Reached 97.34\%. We hope that CogniDot can bring cognitive load and multi-module real-time detection capabilities to the smart health field in a compact form.

\bibliographystyle{ACM-Reference-Format}
\bibliography{Ref/refs}

\appendix

\end{document}